\documentclass[prl,superscriptaddress,twocolumn,floatfix,amsmath,amssymb,showpacs]{revtex4}

\usepackage{graphicx}
\usepackage{dcolumn}
\usepackage{bm}

\begin{document}

\title{Superconductivity induced by phosphorus doping and its coexistence with ferromagnetism in
EuFe$_{2}$(As$_{0.7}$P$_{0.3}$)$_{2}$}

\author{Zhi Ren,$^{1}$ Qian Tao,$^{1}$ Shuai Jiang,$^{1}$ Chunmu Feng,$^{2}$ Cao Wang,$^{1}$ Jianhui Dai,$^{1}$ Guanghan Cao$^{1}$\footnote[1]{Electronic address: ghcao@zju.edu.cn} and Zhu'an Xu\footnote[2]{Electronic address: zhuan@zju.edu.cn}}

\affiliation{Department of Physics, Zhejiang University, Hangzhou 310027, China\\
$^{2}$Test and Analysis Center, Zhejiang University, Hangzhou
310027, China\\}

\date{\today}

\begin{abstract}
We have studied EuFe$_{2}$(As$_{0.7}$P$_{0.3}$)$_{2}$ by the
measurements of x-ray diffraction, electrical resistivity,
thermopower, magnetic susceptibility, magnetoresistance and specific
heat. Partial substitution of As with P results in the shrinkage of
lattice, which generates chemical pressure to the system. It is
found that EuFe$_{2}$(As$_{0.7}$P$_{0.3}$)$_{2}$ undergoes a
superconducting transition at 26 K, followed by ferromagnetic
ordering of Eu$^{2+}$ moments at 20 K. This finding is the first
observation of superconductivity stabilized by internal chemical
pressure, and supplies a rare example showing coexistence of
superconductivity and ferromagnetism in the ferro-arsenide family.
\end{abstract}

\pacs{74.70.Dd; 74.25.Ha; 74.62.Dh; 74.10.+v}

\maketitle

Recently, high-temperature superconductivity has been discovered in
a family of materials containing FeAs
layers.\cite{Kamihara08,Sm&Ce,OV,Wang} The superconductivity is
induced by doping charge carriers into a parent compound,
characterized by an antiferromagnetic (AFM) spin-density-wave (SDW)
transition associated with the FeAs layers.\cite{WNL,Dai} Electron
doping has been realized by the partial substitutions of
F-for-O,\cite{Kamihara08} vacancy-for-O,\cite{OV}
Th-for-Ln,\cite{Wang} Co/Ni-for-Fe,\cite{Co,Co-122,Ni,Ni-122}.
Examples of hole doping include the partial substitutions of
Sr-for-Ln\cite{Wen} and K-for-Ba/Sr/Eu\cite{Rotter,122-Sr,EuK}. All
the chemical doping suppresses the long-range SDW order, eventually
resulting in the emergence of superconductivity. Up to now, no
superconductivity has been reported through doping at the As site.
Apart from carrier doping, application of hydrostatic pressure is
also capable of stabilizing
superconductivity.\cite{P-Ca122,P-SrBa122,P-Eu122}

Among the parent compounds in ferro-arsenide family, EuFe$_2$As$_2$
exhibits peculiar behavior because the moments of Eu$^{2+}$ ions
order antiferromagnetically at relatively high temperature of 20
K.\cite{Eu122,Ren,Jeevan} Magnetoresistance measurements on
EuFe$_2$As$_2$ crystals\cite{Jiang} suggest a strong coupling
between the magnetism of Eu$^{2+}$ ions and conduction electrons,
which may affect or even destroy superconductivity. For example,
though Ni doping in BaFe$_{2}$As$_{2}$ leads to superconductivity up
to 21 K,\cite{Ni-122} ferromagnetism rather than superconductivity
was found in EuFe$_2$As$_2$ by a systematic Ni doping\cite{Ren-Ni}.
Another relevant example is that the superconducting transition
temperature of (Eu,K)Fe$_{2}$As$_{2}$\cite{EuK} is 32 K,
substantially lower than those of (Ba,K)Fe$_{2}$As$_{2}$ ($T_{c}$=38
K)\cite{Rotter} and (Sr,K)Fe$_{2}$As$_{2}$ ($T_{c}$=37
K)\cite{122-Sr}. Resistivity measurement under hydrostatic
pressures\cite{P-Eu122} on EuFe$_{2}$As$_{2}$ crystals showed a
resistivity drop at 29.5 K. However, no zero resistivity could be
achieved, which was ascribed to the AFM ordering of the Eu$^{2+}$
moments\cite{P-Eu122}.

While hetero-valent substitution generally induces charge carriers,
iso-valent substitution may supply chemical pressure. The latter
substitution is of particular interest in EuFe$_2$As$_2$ when As is
partially replaced by P, as suggested theoretically in order to
search for the magnetic quantum criticality without changing the
number of Fe 3$d$-electrons\cite{JDai}. In this Letter, we
demonstrate bulk superconductivity at 26 K in
EuFe$_{2}$(As$_{0.7}$P$_{0.3}$)$_{2}$. For the first time,
superconductivity has been realized through the doping at the As
site in the iron arsenide system. Strikingly, we observe coexistence
of ferromagnetic ordering of Eu$^{2+}$ moments with
superconductivity below 20 K.

Polycrystalline samples of EuFe$_{2}$(As$_{0.7}$P$_{0.3}$)$_{2}$
were synthesized by solid state reaction with EuAs, Fe$_{2}$As and
Fe$_{2}$P. EuAs was presynthesized by reacting Eu grains and As
powders at 873 K for 10 h, then 1073 K for 10 h and finally 1223 K
for another 10 h. Fe$_{2}$As was prepared by reacting Fe powers and
As powders at 873 K for 10 h then 1173 K for 0.5 h. Fe$_{2}$P was
presynthesized by heating Fe powders and P powders very slowly to
873 K and holding for 10 h. Powders of EuAs, Fe$_{2}$As and
Fe$_{2}$P were weighed according to the stoichiometric ratio, ground
and pressed into pellets in an argon-filled glove-box. The pellets
were sealed in evacuated quartz tubes and annealed at 1273 K for 20
h then cooled slowly to room temperature. The phase purity of the
samples was investigated by powder X-ray diffraction, using a
D/Max-rA diffractometer with Cu-K$_{\alpha}$ radiation and a
graphite monochromator. The XRD data were collected with a step-scan
mode in the 2$\theta$ range from 10$^{\circ}$ to 120$^{\circ}$. The
structural refinement was performed using the programme RIETAN
2000.\cite{Izumi}

The electrical resistivity was measured on bar-shaped samples using
a standard four-probe method. The applied current density was $\sim$
0.5 A/cm$^{2}$. The measurements of magnetoresistance, specific
heat, ac magnetic susceptibility and thermopower were performed on a
Quantum Design Physical Property Measurement System (PPMS-9).  DC
magnetic properties were measured on a Quantum Design Magnetic
Property Measurement System (MPMS-5).

Figure 1 show the XRD pattern of
EuFe$_{2}$(As$_{0.7}$P$_{0.3}$)$_{2}$ at room temperature, together
with the profile of the Rietveld refinement using the space group of
I4/\emph{mmm}. No additional diffraction peak is observed. The
refined lattice parameters are \emph{a}=3.889(1){\AA} and
$c$=11.831(3){\AA}. Compared with those of the undoped
EuFe$_{2}$As$_{2}$\cite{Ren}, the \emph{a}-axis is decreased by
0.35\%, the \emph{c}-axis is shortened by 1.8\% and the cell volume
shrinks by 3.2\% for EuFe$_{2}$(As$_{0.7}$P$_{0.3}$)$_{2}$. These
results suggest that the iso-valent substitution of As with P indeed
generates chemical pressure to the system. In addition, the As(P)
position \emph{z} decreases, indicating that the As(P) atoms move
toward the Fe planes. As a consequence, the Fe-As(P)-Fe angle
increases from 110.15$^{\circ}$ to 111.48$^{\circ}$.

\begin{figure}
\includegraphics[width=8cm]{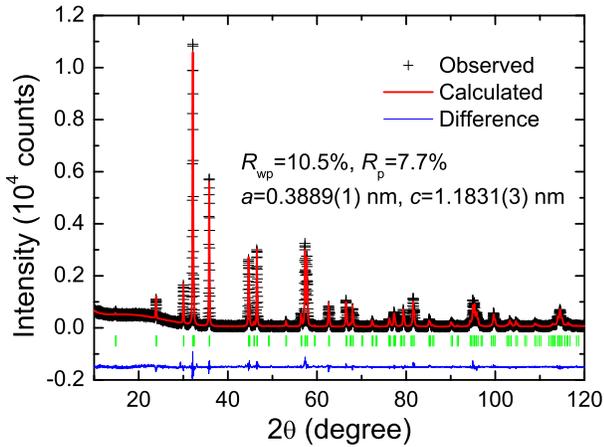}
\caption{(color online). X-ray powder diffraction pattern at room
temperature and the Rietveld refinement profile for
EuFe$_{2}$(As$_{0.7}$P$_{0.3}$)$_{2}$.}
\end{figure}

Figure 2(a) shows the temperature dependence of resistivity ($\rho$)
for EuFe$_{2}$(As$_{0.7}$P$_{0.3}$)$_{2}$ under zero field. The
anomaly associated with the SDW transition in undoped
EuFe$_{2}$As$_{2}$\cite{Ren} is completely suppressed. The
resistivity is linear with temperature down to $\sim$ 90 K and shows
upward deviation from the linearity at lower temperatures. The
resistivity ratio of \emph{R}(300K)/\emph{R}(30K) is 5.2, indicating
high quality of the present sample. Below 29 K, the resistivity
drops steeply, suggesting a superconducting transition. The midpoint
of the transition is 26 K. On closer examination shown in the inset
of Fig. 1(a), however, a small resistivity peak is observed around
16 K, which coincides with the ferromagnetic ordering of Eu$^{2+}$
moments (to be shown below). This observation is reminiscent of the
reentrant superconducting behavior as observed, for example, in
\emph{R}Ni$_{2}$B$_{2}$C$_{2}$ (\emph{R}=Tm, Er, Ho)\cite{Eisaki}.

Figure 2(b) shows the temperature dependence of thermopower ($S$).
The thermopower in the whole temperature range is negative,
indicating that the dominant charge carriers are electron-like.
$\mid$$S$$\mid$ decreases sharply below 29 K, corresponding to the
superconducting transition. However, the $\mid$$S$$\mid$ value does
not drop to zero until $\sim$ 13 K, in relation with the
ferromagnetic ordering of the Eu$^{2+}$ moments.

\begin{figure}
\includegraphics[width=8cm]{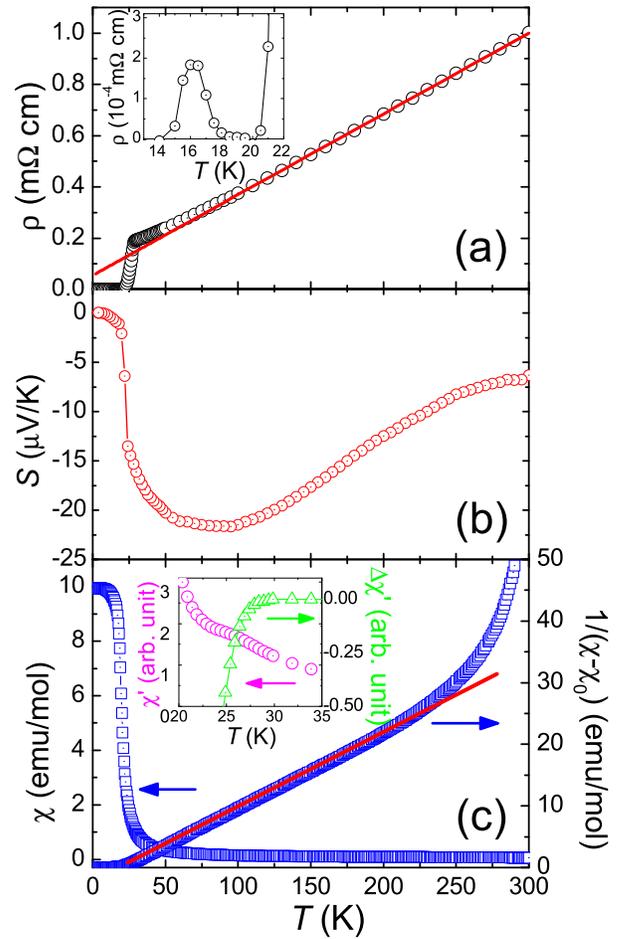}
\caption{(color online). Temperature dependence of resistivity (a),
thermopower (b), and magnetic susceptibility (c) for the
EuFe$_{2}$(As$_{0.7}$P$_{0.3}$)$_{2}$ sample. The inset of (a) shows
an expanded plot. The inset of (c) shows the real part of ac
susceptibility $\chi'$ (left axis) for the same sample. A
diamagnetic signal (right axis) is obtained after subtraction of
paramagnetic contribution of Eu$^{2+}$ moments.}
\end{figure}

In figure 2(c), we show the temperature dependence of field-cooling
dc magnetic susceptibility for EuFe$_{2}$(As$_{0.7}$P$_{0.3}$)$_{2}$
under $\mu_{0}H_{dc}$=0.1 T. The obvious deviation of linearity in
$\chi^{-1}$ above 230 K is probably due to the presence of trace
amount of ferromagnetic Fe$_{2}$P impurity\cite{Cava}. The $\chi$
data between 50 K and 200 K can be well described by the modified
Curie-Weiss law,
\begin{equation}
\chi=\chi_0+\frac{C}{T-\theta},
\end{equation}
where $\chi_0$ denotes the temperature-independent term, $C$ the
Curie-Weiss constant and $\theta$ the Weiss temperature. The fitting
yields \emph{C}= 8.1(1) emu$\cdot$K/mol and $\theta$=22(1) K. The
corresponding effective moment $P_{eff}$=8.0(1) $\mu_{B}$ per
formula unit, close to the theoretical value of 7.94 $\mu_{B}$ for a
free Eu$^{2+}$ ion. $\chi$ increases steeply with decreasing
temperature below 20 K and becomes gradually saturated. Field
dependence of magnetization gives a saturated magnetic moment of
6.9(1)$\mu_{B}$/f.u., as expected for fully paralleled Eu$^{2+}$
moments with \emph{S}=7/2. Therefore, it is concluded that the
moments of Eu$^{2+}$ order ferromagnetically in
EuFe$_{2}$(As$_{0.7}$P$_{0.3}$)$_{2}$, in analogy with
EuFe$_{2}$P$_{2}$\cite{EuFe2P2}. Due to the proximity of
superconducting transition and ferromagnetic ordering, the
superconducting diamagnetic signal is hard to observe unless the
applied magnetic field is very low. As shown in the inset of Fig.
2(c), ac magnetic susceptibility measured under $\mu_{0}H_{ac}$=1 Oe
shows a kink around 26 K. After subtraction of the paramagnetic
contribution from Eu$^{2+}$ moments, a clear diamagnetism is seen,
confirming superconductivity in
EuFe$_{2}$(As$_{0.7}$P$_{0.3}$)$_{2}$.

\begin{figure}
\includegraphics[width=8cm]{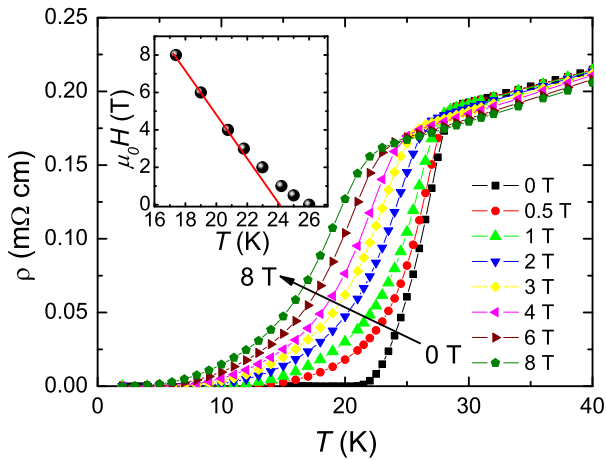}
\caption{(color online). Field dependence of resistive transition
for EuFe$_{2}$(As$_{0.7}$P$_{0.3}$)$_{2}$ sample. The inset shows
the upper critical fields as a function of temperature.}
\end{figure}

The temperature dependence of resistivity under magnetic fields is
shown in figure 3. With increasing magnetic fields, the resistive
transition shifts towards lower temperature and becomes broadened,
further affirming the superconducting transition. The
$T_{c}$(\emph{H}), defined as a temperature where the resistivity
falls to 50\% of the normal state value, is plotted as a function of
magnetic field in the inset of Fig. 3. The $\mu_{0}H_{c2}$-\emph{T}
diagram shows a slight upward curvature, which is probably due to
the multi-band effect\cite{twoband}. The initial slope
$\mu_{0}$$\partial$$H_{c2}$/$\partial$$T$ near $T_{c}$ is -1.18 T/K,
giving an upper critical field of $\mu_{0}H_{c2}$(0) $\sim$ 30 T by
linear extrapolation. It is also noted that the reentrant
superconducting behavior is not obvious under magnetic field, in
contrast with that in \emph{R}Ni$_{2}$B$_{2}$C$_{2}$
superconductors\cite{Eisaki}.

\begin{figure}
\includegraphics[width=8cm]{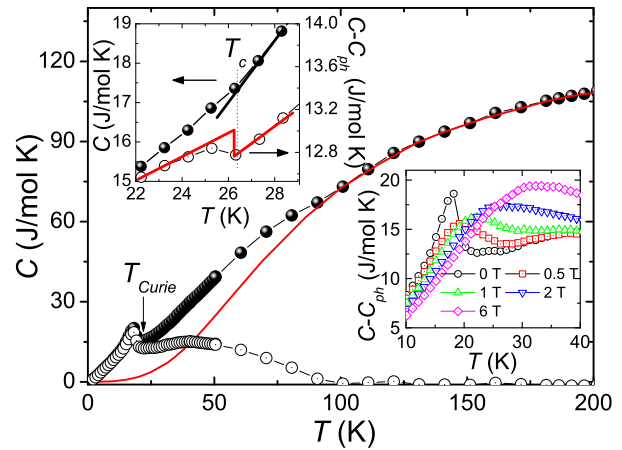}
\caption{(color online). Temperature dependence of specific heat
before (solid symbol) and after (open symbol) deduction of lattice
contribution for the EuFe$_{2}$(As$_{0.7}$P$_{0.3}$)$_{2}$ sample.
The solid curve represents the lattice contribution fitted by the
Debye model. See text for details. The upper left panel shows the
specific heat anomaly around 26 K, which is ascribed to the
superconducting transition. The lower right panel shows the magnetic
specific heat anomaly around 20 K under magnetic fields.}
\end{figure}

Figure 4 shows the specific heat ($C$) for the
EuFe$_{2}$(As$_{0.7}$P$_{0.3}$)$_{2}$ sample. Two anomalies below 30
K are identified. One is a $\lambda$-shape peak with the onset at 20
K, indicating a second-order transition. With increasing magnetic
fields, the anomaly shifts to higher temperatures and becomes
broadened, in accordance with ferromagnetic nature of the
transition. The other anomaly is much weaker but detectable at
$\sim$26 K (shown in the upper-left panel of Fig. 4), which
coincides with the superconducting transition.

To analyze the $C(T)$ data further, it is assumed that the total
specific heat consists of the electronic, phonon and magnetic
components. At high temperatures, the dominant contribution comes
from the phonon component, which can be well described by the Debye
model with only one adjustable parameter (i. e., Debye temperature
$\Theta_{D}$). The data fitting above 100 K gives $\Theta_{D}$ of
345 K. Since $\Theta_{D}$$\sim$1/$\sqrt{M}$ (\emph{M} the molecular
weight), the $\Theta_{D}$ for EuFe$_{2}$(As$_{0.7}$P$_{0.3}$)$_{2}$
is calculated to be 348 K, in good agreement with the fitted value,
using the $\Theta_{D}$ data of 380 K for
EuFe$_{2}$P$_{2}$\cite{EuFe2P2}. The upper bound of electronic
specific heat coefficient is estimated to be 10
mJ/mol$\cdot$K$^{2}$, which is comparable with those of other
iron-arsenide superconductors. The electronic specific heat
contribution is less than 2\% of the total specific heat even at low
temperatures, and thus is not taken into consideration in the
following analysis.

By subtracting the lattice contribution, the specific anomaly due to
the superconducting transition is more prominent. The jump in
specific heat \emph{$\Delta$C}$\approx$300 mJ/mol$\cdot$K at
$T_{c}$, which is of the same order as that in
Ba$_{0.55}$K$_{0.45}$Fe$_{2}$As$_{2}$ crystals\cite{NN}. Meanwhile,
the magnetic entropy associated with the ferromagnetic transition is
16.5 J/mol$\cdot$K, which amounts to 95\% of \emph{R}ln(2\emph{S}+1)
with \emph{S}=7/2 for Eu$^{2+}$ ions. The thermodynamic properties
indicate that the superconducting transition and the ferromagnetic
ordering are both of bulk nature.

A broad specific-heat hump below 90 K is evident after deduction of
the lattice contribution, indicating existence of additional
magnetic contribution. This anomaly is accompanied with the upward
deviation from the linear temperature dependence in resistivity as
shown in Fig. 1(a). Meanwhile, negative magnetoresistance was
observed in the same temperature range, and reaches -6\% at 30 K
under $\mu_{0}H$=8 T (data not shown here). All these features are
probably attributed to the interaction between the moments of
Eu$^{2+}$ ions and conduction electrons. Such interaction may be
responsible for the observed ferromagnetic ordering of Eu$^{2+}$
moments below 20 K.

The isovalent substitution of As with P does not change the number
of Fe 3$d$ electrons, but generates chemical pressure, as manifested
by the shrinkage of lattice. According to a coherent-incoherent
scenario\cite{JDai}, the low energy physics of the FeAs-containing
system is described by the interplay of the coherent excitations
(associated with the itinerant carriers) and incoherent ones
(modeled in terms of Fe localized magnetic moments). The internal
chemical pressure generated via P doping results in the enhancement
of coherent spectral weight, which weakens the SDW ordering and
probably induces a magnetic quantum critical point (QCP). In the
structural point of view, the P doping results in closer distance
between As(P) and Fe planes. As a consequence, the low-energy band
width becomes larger (correspondingly the coherent spectral weight
is enhanced), according to the related band calculations\cite{band}.
Furthermore, the magnetic fluctuations near the QCP may induce
superconductivity, as has been well documented in
literatures.\cite{QCP} This explains the simultaneous suppression of
SDW transition and emergence of superconductivity in
EuFe$_{2}$(As$_{0.7}$P$_{0.3}$)$_{2}$ assuming that the QCP locates
near P content of $\sim$30\%.

The superconducting properties of
EuFe$_{2}$(As$_{0.7}$P$_{0.3}$)$_{2}$ bear similarity with those of
EuFe$_{2}$As$_{2}$ crystals under hydrostatic pressure. As a matter
of fact, the onset temperature of resistive drop is nearly the same
in both cases, suggesting that the effect of internal chemical
pressure is in analogy with application of external physical
pressure. Thus it is expectable to find superconductivity in other
iron arsenide systems via the P-doping strategy.

In summary, we have found superconductivity at 26 K in
EuFe$_{2}$(As$_{0.7}$P$_{0.3}$)$_{2}$. Moreover, ferromagnetic
ordering of Eu$^{2+}$ moments coexists with the superconductivity
below 20 K. The observation of sizable anomalies in thermodynamic
properties, concomitant with the transitions, indicates that both of
them are bulk phenomena. The interplay of superconductivity and
ferromagnetism may bring about exotic properties and provide clues
to the superconductivity mechanism, which renders
EuFe$_{2}$(As$_{1-x}$P$_{x}$)$_{2}$ worthy of further exploration.

\begin{acknowledgments}
This work is supported by the NSF of China, National Basic Research
Program of China (No. 2007CB925001) and the PCSIRT of the Ministry
of Education of China (IRT0754).
\end{acknowledgments}


\begin{thebibliography}{00}
\bibitem{Kamihara08}Y. Kamihara \emph{et al.}, J. Am. Chem. Soc. \textbf{130}, 3296 (2008).
\bibitem{Sm&Ce}X. H. Chen \emph{et al.}, Nature \textbf{453}, 761 (2008); G. F. Chen \emph{et al.}, Phys. Rev. Lett. \textbf{100}, 247002 (2008).
\bibitem{OV}Z. A. Ren \emph{et al.}, Europhysics Lett. \textbf{83}, 17002 (2008); H. Kito, H. Eisaki and A. Iyo, J. Phys. Soc. Jpn. \textbf{77},
063707 (2008).
\bibitem{Wang}C. Wang \emph{et al.}, Europhysics Lett. \textbf{83}, 67006 (2008).
\bibitem{WNL}J. Dong \emph{et al.}, Europhysics Lett. \textbf{83}, 27006 (2008).
\bibitem{Dai}C. Cruz \emph{et al.}, Nature \textbf{453}, 899 (2008).
\bibitem{Co}A. S. Sefat \emph{et al.}, Phys. Rev. B \textbf{78}, 104505
(2008); G. H. Cao \emph{et al.}, arXiv: 0807.1304 (2008).
\bibitem{Co-122}A. S. Sefat \emph{et al.}, Phys. Rev. Lett. \textbf{101},
117004 (2008); A. Leithe-Jasper, W. Schnelle, C. Geibel, and H.
Rosner, Phys. Rev. Lett. \textbf{101}, 207004 (2008).
\bibitem{Ni}G. H. Cao \emph{et al.}, arXiv:0807.4328 (2008).
\bibitem{Ni-122}J. L. Li \emph{et al.}, New J. Phys. \emph{in
press}.
\bibitem{Wen}H. H. Wen \emph{et al.}, Europhysics Lett. \textbf{82}, 17009 (2008).
\bibitem{Rotter}M. Rotter, M. Tegel, and D. Johrendt, Phys. Rev. Lett. \textbf{101},
107006 (2008).
\bibitem{122-Sr}K. Sasmal \emph{et al.}, Phys. Rev. Lett. \textbf{101}, 107007
(2008); G. F. Chen \emph{et al.}, Chin. Phys. Lett. \textbf{25},
3403 (2008).
\bibitem{EuK} H. S. Jeevan \emph{et al.},  Phys. Rev. B, \textbf{78}, 092406 (2008).

\bibitem{P-Ca122}M. S. Torikachvili, S. L. Budko, N. Ni, and P. C. Canfield, Phys. Rev. Lett. \textbf{101}, 057006 (2008); T. Park \emph{et
al.}, J. Phys.: Condensed Matter \textbf{20}, 322204 (2008).
\bibitem{P-SrBa122}P. L. Alireza \emph{et al.}, J. Phys.: Condensed Matter \textbf{21}, 012208 (2008).
\bibitem{P-Eu122}C. F. Miclea \emph{et al.}, arXiv:0808.2026 (2008).

\bibitem{Eu122}R. Marchand, W. Jeitschko, J. Solid State Chem. \textbf{24}, 351(1978).
\bibitem{Ren}Z. Ren \emph{et al.},  Phys. Rev. B \textbf{78}, 052501 (2008).
\bibitem{Jeevan}H. S. Jeevan \emph{et al.}, Phys. Rev. B, \textbf{78}, 052502 (2008).
\bibitem{Jiang}S. Jiang \emph{et al.}, New J. Phys. \emph{in press}.

\bibitem{Ren-Ni}Z. Ren \emph{et al.}, arXiv:0810.2595 (2008).

\bibitem{JDai}J. H. Dai \emph{et al.}, arXiv:0808.0305 (2008).
\bibitem{Izumi}F. Izumi \emph{et al.}, Mater. Sci. Forum. \textbf{198}, 321 (2000).
\bibitem{Eisaki}H. Eisaki \emph{et al.}, Phys. Rev. B (R), \textbf{50}, 647 (1994).
\bibitem{Cava}T. M. McQueen \emph{et al.}, Phys. Rev. B, \textbf{78}, 024521 (2008).
\bibitem{EuFe2P2}H. Raffius \emph{et al.}, J. Phys. Chem. Solids, \textbf{52}, 787 (1991).
\bibitem{twoband}F. Hunte \emph{et al.}, Nature \textbf{453}, 903 (2008).
\bibitem{EuK1}Anupam \emph{et al.}, arXiv:0812.1131 (2008).
\bibitem{NN}N. Ni \emph{et al.}, Phys. Rev. B, \textbf{78}, 014507 (2008).
\bibitem{band}V. Vildosola, L. Pourovskii, R. Arita, S. Biermann, and A. Georges, Phys. Rev. B, \textbf{78}, 064518 (2008).
\bibitem{QCP}P. Gegenwart, Q. Si, and F. Steglich, Nature Phys. \textbf{4},
186 (2008).

\end{thebibliography}
\end{document}